# 3D solution of the full Faddeev equations for the Nd break-up reaction without 3NF effect

Reza Ramazani-Sharifabadi,[1,†,*]  and Iman Ziaeian [2]

[1]Physics Department, Faculty of Science, Imam Khomeini International University, P.O. Box 34149-16818, Qazvin, Iran
[3]Physics and Accelerators Research School, Nuclear Science and Technology Research Institute, AEOI, P.O. Box 1439951113, Tehran, Iran

**ABSTRACT**. A recently developed three-dimensional formalism for the nucleon-deuteron breakup channel initially considered only the leading-order term of the Faddeev equations, using the nucleon-nucleon T-matrix to compute the breakup amplitude. In the present study, we extend that formalism by solving the full three-nucleon Faddeev equation without three-nucleon-force contributions in a three-dimensional approach in which momentum vectors are used directly as variables. This formalism is well suited for projectile energies in scattering processes above the pion-production threshold, where partial-wave expansions become inefficient. To treat the moving singularities of the free three-nucleon propagator, we introduce a new method that treats these singularities in a manner analogous to the simple poles appearing in the Lippmann-Schwinger equation. This approach evaluates the moving singularities directly through the center-of-mass energy of the 23-subsystem and eliminates the need to partition the q-domain into intervals. The resulting formulation opens a pathway toward a systematic investigation of the complex singularities in few-body scattering processes within the Faddeev framework.

## I. INTRODUCTION.

It is common to interpret the interactions between nucleons in terms of meson exchange [1]. Several computational techniques have been developed, benefiting from recent advances in high-performance computing and numerical methods, to study few-body interactions primarily three- and four-nucleon systems [2-4]. Exact predictions can now be obtained using high-precision (semi)phenomenological nucleon-nucleon potentials [5-7] as well as those derived from the chiral effective field theory [8-9].

One of the most powerful and widely used approaches for the exact numerical treatment of few-body systems is the Faddeev formalism [10]. The Faddeev equations, originally formulated for three-body systems, have been extended in various ways to address four-body and larger systems. The Faddeev-Yakubovsky (FY) formalism [11] and the Alt-Grassberger-Sandhas (AGS) equations [12] are the standard tools for solving such few-body scattering problems. In the Faddeev approach, the wave function of the system is decomposed into components that satisfy a coupled set of integral equations, which can be solved using modern computational techniques.

A conventional method for determining scattering observables in few-nucleon scattering involves solving the Faddeev equations in either momentum [13] or coordinate [14] space. This approach exploits rotational symmetry by employing a partial-wave basis [2]. This technique is well-established and, at energies below the pion-production threshold, provides a transparent physical interpretation. However, as the energy increases, the number of partial waves required for convergence grows rapidly, making alternative methods that solve the scattering equations directly in terms of vector variables increasingly important. This is where the three-dimensional (3D) formalism comes into play [15-17]. The 3D approach eliminates the need for angular momentum decomposition and provides a direct solution of the Faddeev equations in full three-dimensional momentum space, without relying on approximations inherent in partial-wave expansions. The 3D formulation of the Faddeev equations has recently been applied to three- [18-20] and four-body bound states [21], as well as to three-nucleon (3N) scattering processes [22-23].

Several techniques have been developed to describe NN and 3N scattering processes without relying on partial-wave bases. One such approach, based on a momentum-helicity basis tied to the total NN spin, was first introduced in [24] to describe NN scattering in a three-dimensional framework and was later extended to three-boson systems [18-25] as well as three-nucleon bound-state calculations [26]. Three-dimensional formulations have also recently been applied to pion-nucleons scattering [27] and the proton scattering on light nuclei [28].

In the momentum-helicity method, the scattering amplitude is expressed in terms of the helicity states of the interacting particles. An important advantage of this approach is that these states are eigenstates of the helicity operator appearing in the NN potential, which simplifies the structure of the equations. Unlike spin, the helicity is invariant under the rotation which significantly simplifies the analysis of rotational invariance. Another benefit is that it facilitates the inclusion of the relativistic effects in the calculation [29].

Over recent decades, high-precision experimental programs have produced an extensive body of three- and four-nucleon scattering data, enabling detailed studies of nuclear interaction [30-40]. In particular, the available Nd scattering data has provided invaluable insights not only into the spin and momentum dependence of nuclear forces but also into the reaction mechanisms underlying multiple re-scattering processes. However, comparisons between experimental results and theoretical predictions reveal discrepancies that exceed systematic uncertainties, especially for certain spin observables, indicating that the important features of the NN interaction and three-nucleon force (3NF) effects are still not fully understood [41-44].

Three-nucleon scattering at intermediate and high energies (a few hundred MeV) exhibit enhanced sensitivity to short-distance structure of 3N and NN forces. Nd scattering, as one of the simplest few-nucleon systems, with its extensive experimental database and refined theoretical treatments based on the momentum-space Faddeev formalism, provides a uniquely valuable laboratory for exploring these effects.

Nucleon-deuteron break-up process was first evaluated within the Faddeev scheme in [45-46] where only the leading term of the multiple scattering series was considered. Comparisons between calculations based on the leading term of the Faddeev equations and full partial-wave Faddeev calculations indicates that re-scattering contributions at intermediate energies play a significant role in observables such as cross sections and analyzing powers [45]. Therefore, the aim of the present study is to extend this momentum-space formalism by solve full Faddeev equations. Although three-nucleon-force effects are omitted for clarity, the full treatment still yields a significantly improved description of the underlying reaction dynamics and provides new insights into Nd scattering and breakup processes. The resulting formulation also provides a robust basis for a systematic investigation of complex singularities in few-body scattering processes within the Faddeev framework.

## II. FADDEEV EQUATIONS FOR THREE-NUCLEON SCATTERING

*Contact author: r.ramazani@sci.ikiu.ac.ir

†Contact author: reza_ramazani@ut.ac.ir

The full Nd break-up operator in the Faddeev framework, $U^{full}$, is defined as

$$U^{full} = (1 + P)T_F, \qquad (1)$$

Where $T_F$ is the Faddeev transition operator which satisfies the Faddeev equation describing the breakup process of three identical nucleons,

$$T_F = TP + TPG_0T_F + (1 + TG_0)V_4^{(1)}(1 + P) + (1 + TG_0)V_4^{(1)}G_0(1 + P)T_F. \qquad (2)$$

The free three-nucleon propagator is given by

$$G_0(E) = G_0\left(\frac{3q_0^2}{4m} + E_d\right) = \left(\frac{3q_0^2}{4m} + E_d - H_0\right)^{-1}, \qquad (3)$$

where E is the total energy of the system and $E_d$ is the binding energy of the deuteron. The permutation operator is defined as

$$P = P_{12}P_{23} + P_{13}P_{23}. \qquad (4)$$

By excluding three-nucleon force (3NF) contributions, the Faddeev operator reduces to

$$T_F = TP + TPG_0T_F. \qquad (5)$$

The matrix elements of the full Nd breakup amplitude are defined as

$$U^{full}(\vec{p},\vec{q},\vec{q_0}) = <\vec{p}\vec{q}, m_{s1}m_{s2}m_{s3}\tau_1\tau_2\tau_3|(1 + P)T_F|\vec{q_0}, m_{s1}^0\tau_1^0\psi_d^{M_d}>, \qquad (6)$$

where the final free three-nucleon state is defined as

$$|\vec{p}\vec{q}, m_{s1}m_{s2}m_{s3}\tau_1\tau_2\tau_3> \equiv |\vec{q}, m_{s1}\tau_1>|\vec{p}, m_{s2}m_{s3}\tau_2\tau_3> \qquad (7)$$

in which $m_i$ and $\tau_i$ (i=1,2,3) denote the spin and isospin projections of the nucleons. The deuteron state, $|\psi_d^{M_d}>$, is the only anti-symmetrized state in the initial channel. Thus, the initial state is written as

$$|\vec{q_0}, m_{s1}^0\tau_1^0\psi_d^{M_d}> \equiv |\vec{q_0}, m_{s1}^0\tau_1^0>|\psi_d^{M_d}>. \qquad (8)$$

Here, $m_{s1}^0$ and $\tau_1^0$ denote the spin and isospin projection of the projectile nucleon and, $M_d$ denotes the projection of the total angular momentum of the deuteron along an arbitrary z axis. The $\vec{p}$ and $\vec{q}$ are Jacobi momenta used to describe the three-nucleon kinematics in the final sate

$$\vec{p} = \frac{1}{2}(\vec{k}_2 - \vec{k}_3), \qquad (9.a)$$
$$\vec{q} = \frac{2}{3}\left[\vec{k}_1 - \frac{1}{2}(\vec{k}_2 - \vec{k}_3)\right], \qquad (9.b)$$

where $\vec{k}_i$'s (i=1,2,3) indicate the laboratory momenta of three nucleons. $\vec{q_0}$ is also the relative momentum of the projectile to the deuteron target. Without loss of generality, nucleon 1 is designate as the projectile, while nucleons 2 and 3 form the deuteron state as the two-nucleon subsystem in the initial state. The full Nd break-up amplitude, $U^{full}(\vec{p},\vec{q},\vec{q_0})$, consists of three components

$U^{(1)}(\vec{p},\vec{q},\vec{q_0}) = <\vec{p}\vec{q}, m_{s1}m_{s2}m_{s3}\tau_1\tau_2\tau_3|T_F|\vec{q_0}, m_{s1}^0\tau_1^0\psi_d^{M_d}>$ (10)

$U^{(2)}(\vec{p},\vec{q},\vec{q_0}) = <\vec{p}\vec{q}, m_{s1}m_{s2}m_{s3}\tau_1\tau_2\tau_3|P_{12}P_{23}T_F|\vec{q_0}, m_{s1}^0\tau_1^0\psi_d^{M_d}>$ (11)

$U^{(3)}(\vec{p},\vec{q},\vec{q_0}) = <\vec{p}\vec{q}, m_{s1}m_{s2}m_{s3}\tau_1\tau_2\tau_3|P_{13}P_{23}T_F|\vec{q_0}, m_{s1}^0\tau_1^0\psi_d^{M_d}>$ (12)

Each amplitude $U^{(i)}(\vec{p},\vec{q},\vec{q_0})$ corresponds to the case in which the $i$-th nucleon is the free particle while the remaining two nucleons ($j$ and $k$), form the interacting two-nucleon subsystem. It has been shown that $U^{(2)}(\vec{p},\vec{q},\vec{q_0})$ and $U^{(3)}(\vec{p},\vec{q},\vec{q_0})$ differ from $U^{(1)}(\vec{p},\vec{q},\vec{q_0})$ only by their variables. Therefore, it is sufficient to derive an explicit expression for a single component. The other two amplitudes $U^{(2)}(\vec{p},\vec{q},\vec{q_0})$ and $U^{(3)}(\vec{p},\vec{q},\vec{q_0})$ are obtained as following

$U^{(2)}(\vec{p},\vec{q},\vec{q_0}) = <(-\frac{1}{2}\vec{p}-\frac{3}{4}\vec{q})(\vec{p}-\frac{1}{2}\vec{q}), m_{s2}m_{s3}m_{s1}\tau_2\tau_3\tau_1|T_F|\vec{q_0}, m_{s1}^0\tau_1^0\psi_d^{M_d}>$ (13)

$U^{(3)}(\vec{p},\vec{q},\vec{q_0}) = <(-\frac{1}{2}\vec{p}+\frac{3}{4}\vec{q})(-\vec{p}-\frac{1}{2}\vec{q}), m_{s3}m_{s1}m_{s2}\tau_3\tau_1\tau_2|T_F|\vec{q_0}, m_{s1}^0\tau_1^0\psi_d^{M_d}>$ (14)

We select $U^{(1)}(\vec{p},\vec{q},\vec{q_0})$ for further calculations. Thus, the Nd break-up amplitude for the case where nucleon 1 is the free nucleon is written as

$U^{(1)}(\vec{p},\vec{q},\vec{q_0}) = T_F(\vec{p},\vec{q},\vec{q_0}) = U_0^{(1)}(\vec{p},\vec{q},\vec{q_0}) + U_1^{(1)}(\vec{p},\vec{q},\vec{q_0})$ (15)

where $U^{(1)}(\vec{p},\vec{q},\vec{q_0}) = T_F(\vec{p},\vec{q},\vec{q_0})$ is defined in Eq. 10 and the two contributions, $U_i^{(1)}(\vec{p},\vec{q},\vec{q_0})$ ($i=0,1$), are given by

$U_0^{(1)}(\vec{p},\vec{q},\vec{q_0}) = <\vec{p}\vec{q}, m_{s1}m_{s2}m_{s3}\tau_1\tau_2\tau_3|TP|\vec{q_0}, m_{s1}^0\tau_1^0\psi_d^{M_d}>$ (16)

$U_1^{(1)}(\vec{p},\vec{q},\vec{q_0}) = <\vec{p}\vec{q}, m_{s1}m_{s2}m_{s3}\tau_1\tau_2\tau_3|TPG_0T_F|\vec{q_0}, m_{s1}^0\tau_1^0\psi_d^{M_d}>$ (17)

**A. THE FIRST TERM**

The first-order (leading) contribution to the three-nucleon Faddeev equation has been formulated using momentum vectors as variables within the helicity representation. The leading term of the nucleon-deuteron breakup amplitude $U_0^{(1)}(\vec{p},\vec{q},\vec{q_0})$, given in Eq. (16), was evaluated in [45] as

$U_0^{(1)}(\vec{p},\vec{q},\vec{q_0}) = \sum_{m_s'} \frac{(-)^{\frac{1}{2}+\tau_1}}{4\sqrt{2}} \delta_{\tau_2+\tau_3,\tau_1^0-\tau_1}$

$e^{-i(\Lambda_0\varphi_p-\Lambda_0'\varphi_{\pi_0})} C\left(\frac{1}{2}\frac{1}{2}1;m_s'm_{s1}\right)$

$\sum_l C(l11;M_d-m_s'-m_{s1},m_s'+m_{s1})$

$Y_{l,M_d-m_s'-m_{s1}}\left(\hat{\vec{\pi}}_0'\right)\psi_l(\pi_0')$

$\sum_{S\Pi t}(1-\eta_\Pi(-)^{S+t})\, C\left(\frac{1}{2}\frac{1}{2}t;\tau_2\,\tau_3\right)$

$C\left(\frac{1}{2}\frac{1}{2}t;\tau_1^0,-\tau_1\right) C\left(\frac{1}{2}\frac{1}{2}S;m_{s2}m_{s3}\Lambda_0\right) C\left(\frac{1}{2}\frac{1}{2}S;m_{s1}^0m_s'\Lambda_0'\right)$

$\sum_{\Lambda\Lambda'} d^S_{\Lambda_0\Lambda}(\theta_p) d^S_{\Lambda_0'\Lambda'}(\theta_{\pi_0})$

$\sum_{N=-S}^{S} \frac{e^{-iN(\varphi_p-\varphi_{\pi_0})} d^S_{N\Lambda}(\theta_p) d^S_{N\Lambda'}(\theta_{\pi_0})}{d^S_{\Lambda'\Lambda}(\theta')} T^{\pi St}_{\Lambda\Lambda'}(p,\pi_0,\cos\theta_0';E_p)$,, (18.a)

with

$\vec{\pi}_0 \equiv \frac{\vec{q}}{2}+\vec{q_0},\ \vec{\pi}_0' \equiv -\vec{q}-\frac{\vec{q_0}}{2},\ \vec{\pi}_0'' = -\frac{\vec{\pi}_0}{2}-\frac{3}{4}\vec{q_0},$ (18.b)

and

$\cos\theta_0' = \cos\theta_p\cos\theta_{\pi_0} + \sin\theta_p\sin\theta_{\pi_0}\cos(\varphi_p-\varphi_{\pi_0}).$ (18.c)

**B. THE SECOND TERM**

Higher-order re-scattering processes, encapsulated in the second term of Faddeev equation, $U_1^{(1)}(\vec{p},\vec{q},\vec{q_0})$, contribute significantly to the cross section and certain spin observables, particularly at higher projectile energies [45]. Furthermore, a complete solution of the Faddeev equations necessarily requires the evaluation of $U_1^{(1)}(\vec{p},\vec{q},\vec{q_0})$ given in Eq. (17). To compute this term, we insert the standard completeness relation

$\sum_{m_{s1}m_{s2}m_{s3}\tau_1\tau_2\tau_3} \int d\vec{p} \int d\vec{q} \, |\vec{p}\vec{q}, m_{s1}m_{s2}m_{s3}\tau_1\tau_2\tau_3 > < \vec{p}\vec{q}, m_{s1}m_{s2}m_{s3}\tau_1\tau_2\tau_3| = 1,$ (19)

three times into the Eq. (17). This leads to

$U_1^{(1)}(\vec{p},\vec{q},\vec{q_0}) = \sum_{m_{s1}'m_{s2}'m_{s3}'\tau_1'\tau_2'\tau_3'} \sum_{m_{s1}''m_{s2}''m_{s3}''\tau_1''\tau_2''\tau_3''}$

*Contact author: r.ramazani@sci.ikiu.ac.ir
†Contact author: reza_ramazani@ut.ac.ir

$$\sum_{m_{s1}''' m_{s2}''' m_{s3}''' \tau_1''' \tau_2''' \tau_3'''} \int d\vec{p}'$$

$$\int d\vec{p}'' \int d\vec{p}''' \int d\vec{q}' \int d\vec{q}'' \int d\vec{q}'''$$

$$<\vec{p}\vec{q}, m_{s1}m_{s2}m_{s3}\tau_1\tau_2\tau_3|T|\vec{p}'\vec{q}'m_{s1}'m_{s2}'m_{s3}'\tau_1'\tau_2'\tau_3'><$$
$$\vec{p}'\vec{q}'m_{s1}'m_{s2}'m_{s3}'\tau_1'\tau_2'\tau_3'|P|\vec{p}''\vec{q}''m_{s1}''m_{s2}''m_{s3}''\tau_1''\tau_2''\tau_3''>$$
$$<\vec{p}''\vec{q}''m_{s1}''m_{s2}''m_{s3}''\tau_1''\tau_2''\tau_3''|G_0$$
$$|\vec{p}'''\vec{q}'''m_{s1}'''m_{s2}'''m_{s3}'''\tau_1'''\tau_2'''\tau_3'''><$$
$$\vec{p}'''\vec{q}'''m_{s1}'''m_{s2}'''m_{s3}'''\tau_1'''\tau_2'''\tau_3'''|T_F|\overrightarrow{q_0}, m_{s1}^0 \tau_1^0 \psi_d^{M_d}>$$
(20)

Applying the free propagator $G_0$ to the corresponding state, the Eq. (20) takes the form

$$U_1^{(1)}(\vec{p},\vec{q},\overrightarrow{q_0}) = \sum_{m_{s1}'m_{s2}'m_{s3}'\tau_1'\tau_2'\tau_3'} \sum_{m_{s1}''m_{s2}''m_{s3}''\tau_1''\tau_2''\tau_3''}$$
$$\int d\vec{p}' \int d\vec{p}'' \int d\vec{q}' \int d\vec{q}'' \, G_0^* <$$
$$\vec{p}\vec{q}, m_{s1}m_{s2}m_{s3}\tau_1\tau_2\tau_3|T|\vec{p}'\vec{q}'m_{s1}'m_{s2}'m_{s3}'\tau_1'\tau_2'\tau_3'><$$
$$\vec{p}'\vec{q}'m_{s1}'m_{s2}'m_{s3}'\tau_1'\tau_2'\tau_3'|P|\vec{p}''\vec{q}''m_{s1}''m_{s2}''m_{s3}''\tau_1''\tau_2''\tau_3''>$$
$$<\vec{p}''\vec{q}''m_{s1}''m_{s2}''m_{s3}''\tau_1''\tau_2''\tau_3''|T_F|\overrightarrow{q_0}, m_{s1}^0 \tau_1^0 \psi_d^{M_d}>,$$
(21)

Where the propagator is

$$G_0^* = (\frac{3}{4m}(q_0^2 - q''^2) + E_d + i\epsilon - \frac{p''^2}{m})^{-1}. \quad (22)$$

Using the momentum-space properties of the initial state and the fact that operator $T$ acts solely within the interacting nucleon pairs, Eq. (21) can be re-expressed as

$$U_1^{(1)}(\vec{p},\vec{q},\overrightarrow{q_0}) = \sum_{m_{s1}'m_{s2}'m_{s3}'\tau_1'\tau_2'\tau_3'} \sum_{m_{s1}''m_{s2}''m_{s3}''\tau_1''\tau_2''\tau_3''}$$
$$\int d\vec{p}' \int d\vec{p}'' \int d\vec{q}' \int d\vec{q}'' \, G_0^*$$
$$<\vec{q}m_{s1}\tau_1|\vec{q}'m_{s1}'\tau_1'>$$
$$<\vec{p}, m_{s2}m_{s3}\tau_2\tau_3|T|\vec{p}'m_{s2}'m_{s3}'\tau_2'\tau_3'>$$
$$<\vec{p}'\vec{q}'m_{s1}'m_{s2}'m_{s3}'\tau_1'\tau_2'\tau_3'|P|\vec{p}''\vec{q}''m_{s1}''m_{s2}''m_{s3}''\tau_1''\tau_2''\tau_3''>$$
$$<\vec{p}''\vec{q}''m_{s1}''m_{s2}''m_{s3}''\tau_1''\tau_2''\tau_3''|T_F|\overrightarrow{q_0}, m_{s1}^0 \tau_1^0 \psi_d^{M_d}>.$$
(23)

Applying the resulting δ-functions, we obtain

$$U_1^{(1)}(\vec{p},\vec{q},\overrightarrow{q_0}) = \sum_{m_{s2}'m_{s3}'\tau_2'\tau_3'} \sum_{m_{s1}''m_{s2}''m_{s3}''\tau_1''\tau_2''\tau_3''}$$
$$\int d\vec{p}' \int d\vec{p}'' \int d\vec{q}'' \, G_0^* <$$
$$\vec{p}, m_{s2}m_{s3}\tau_2\tau_3|T(E_p)|\vec{p}'m_{s2}'m_{s3}'\tau_2'\tau_3'><$$
$$\vec{p}'\vec{q}m_{s1}m_{s2}'m_{s3}'\tau_1\tau_2'\tau_3'|P|\vec{p}''\vec{q}''m_{s1}''m_{s2}''m_{s3}''\tau_1''\tau_2''\tau_3''>$$
$$<\vec{p}''\vec{q}''m_{s1}''m_{s2}''m_{s3}''\tau_1''\tau_2''\tau_3''|T_F|\overrightarrow{q_0}, m_{s1}^0 \tau_1^0 \psi_d^{M_d}>.$$
(24)

*Contact author: r.ramazani@sci.ikiu.ac.ir

†Contact author: reza_ramazani@ut.ac.ir

The nucleon-nucleon T matrix is evaluated at the center of mass (c.m.) energy $E_p$ of the 23-subsystem,

$$E_p = \frac{p^2}{m} = \frac{3}{4m}(q_0^2 - q^2) + E_d. \quad (25)$$

To streamline the treatment of the moving logarithmic singularities, we rewrite the propagator in terms of the c.m. energy of the 23-subsystem, obtaining

$$G_0^* = (\frac{p^2}{m} + i\epsilon + \frac{3}{4m}q^2 - \frac{3}{4m}q''^2 - \frac{p''^2}{m})^{-1}. \quad (26)$$

The evaluation of the permutation operator in Eq. (24) yields

$$<\vec{p}'\vec{q}m_{s1}m_{s2}'m_{s3}'\tau_1\tau_2'\tau_3'|P|\vec{p}''\vec{q}''m_{s1}''m_{s2}''m_{s3}''\tau_1''\tau_2''\tau_3''>$$
$$=<\vec{p}'\vec{q}|(P_{13}P_{23} + P_{12}P_{23})|\vec{p}''\vec{q}''>$$
$$<m_{s1}m_{s2}'m_{s3}'\tau_1\tau_2'\tau_3'|(P_{13}P_{23}$$
$$+ P_{12}P_{23})|m_{s1}''m_{s2}''m_{s3}''\tau_1''\tau_2''\tau_3''>$$
$$= \delta(\vec{p}''-\pi')\delta(\vec{p}'$$
$$-\pi)\delta_{m_{s1}m_{s3}''}\delta_{m_{s2}'m_{s1}''}\delta_{m_{s3}'m_{s2}''}\delta_{\tau_1\tau_3''}\delta_{\tau_2'\tau_1''}\delta_{\tau_3'\tau_2''}$$
$$+ \delta(\vec{p}''+\pi')\delta(\vec{p}'+$$
$$\pi)\delta_{m_{s1}m_{s2}''}\delta_{m_{s2}'m_{s3}''}\delta_{m_{s3}'m_{s1}''}\delta_{\tau_1\tau_2''}\delta_{\tau_2'\tau_3''}\delta_{\tau_3'\tau_1''}, \quad (27)$$

where

$$\vec{\pi} \equiv \frac{\vec{q}}{2} + \vec{q}'', \vec{\pi}' \equiv -\vec{q} - \frac{\vec{q}''}{2}, \vec{\pi}' = -\frac{\vec{\pi}}{2} - \frac{3}{4}\vec{q}. \quad (28)$$

Eq. (27) consists of two terms. Applying the corresponding δ functions from the first term of Eq. (27) into Eq. (24) yields

$$U_1^{(1)}(\vec{p},\vec{q},\overrightarrow{q_0}) = \sum_{m_{s2}'m_{s3}'\tau_2'\tau_3'} \int d\vec{q}''$$
$$G_0^* <\vec{p}, m_{s2}m_{s3}\tau_2\tau_3|T(E_p)|\vec{\pi}, m_{s2}'m_{s3}'\tau_2'\tau_3'>$$
$$<\vec{\pi}', m_{s3}'m_{s1}\tau_3'\tau_1|<\vec{q}'', m_{s2}'\tau_2'|T_F|\overrightarrow{q_0}, m_{s1}^0 \tau_1^0>$$
$$|\psi_d^{M_d}>. \quad (29.a)$$

Similarly, applying corresponding δ functions from the second term gives

$$U_1^{(1)}(\vec{p},\vec{q},\overrightarrow{q_0}) = \sum_{m_{s2}'m_{s3}'\tau_2'\tau_3'} \int d\vec{q}''$$
$$G_0^* <\vec{p}, m_{s2}m_{s3}\tau_2\tau_3|T(E_p)|(-\vec{\pi}), m_{s2}'m_{s3}'\tau_2'\tau_3'>$$
$$<(-\vec{\pi}')m_{s1}m_{s2}'\tau_1\tau_2'|<\vec{q}''m_{s3}'\tau_3'|T_F|\overrightarrow{q_0}m_{s1}^0\tau_1^0>$$
$$|\psi_d^{M_d}>. \quad (29.b)$$

Using the δ functions $\delta(\vec{p}''-\pi')$ and $\delta(\vec{p}''+\pi')$ from Eq. (27), the propagator $G_0^*$ is rewritten based on the defined variables, $\vec{\pi}$ and $\vec{\pi}'$, in Eq. (28) as

$$G_0^* = \left(\frac{p^2}{m} + i\epsilon + \frac{3}{4m}q^2 - \frac{3}{4m}q''^2 - \frac{\pi'^2}{m}\right)^{-1} =$$

$$\left(\frac{p^2}{m} + i\epsilon + \frac{3}{4m}q^2 - \frac{3}{4m}q''^2 - \frac{\left(-\vec{q}-\frac{\vec{q}''}{2}\right)^2}{m}\right)^{-1} =$$

$$\left(\frac{p^2}{m} + i\epsilon + \frac{3}{4m}q^2 - \frac{3}{4m}q''^2 - \frac{1}{m}q^2 - \frac{1}{4m}q''^2 - \frac{\vec{q}\cdot\vec{q}''}{m}\right)^{-1} = \left(\frac{p^2}{m} + i\epsilon - \frac{\left(\frac{\vec{q}}{2}+\vec{q}''\right)^2}{m}\right)^{-1} = \left(\frac{p^2}{m} + i\epsilon - \frac{\pi^2}{m}\right)^{-1}. \quad (30)$$

Considering the anti-symmetry property of the deuteron state, $|\psi_d^{M_d}\rangle$, the two terms in Eqs. (29) can be merged through the action of the permutation operator $P_{23}$:

$$U_1^{(1)}(\vec{p},\vec{q},\vec{q_0}) = \sum_{m'_{s2}m'_{s3}\tau'_2\tau'_3} \int d\vec{q}''$$
$$G_0^* \langle \vec{p}, m_{s2}m_{s3}\tau_2\tau_3|T(E_p)(1-P_{23})|\vec{\pi}, m'_{s2}m'_{s3}\tau'_2\tau'_3\rangle$$
$$\times \langle \vec{\pi}', m'_{s3}m_{s1}\tau'_3\tau_1|\langle \vec{q}'', m'_{s2}\tau'_2|T_F|\vec{q_0}, m_{s1}^0\tau_1^0\rangle$$
$$|\psi_d^{M_d}\rangle$$
$$= \sum_{m'_{s2}m'_{s3}\tau'_2\tau'_3} \int d\vec{q}''$$
$$G_0^* \,_a\langle \vec{p}, m_{s2}m_{s3}\tau_2\tau_3|T(E_p)|\vec{\pi}, m'_{s2}m'_{s3}\tau'_2\tau'_3\rangle$$
$$\langle \vec{\pi}', m'_{s3}m_{s1}\tau'_3\tau_1|\langle \vec{q}'', m'_{s2}\tau'_2|T_F|\vec{q_0}, m_{s1}^0\tau_1^0\rangle$$
$$|\psi_d^{M_d}\rangle \quad (31)$$

In the last equality, the matrix element

$$_a\langle \vec{p}, m_{s2}m_{s3}\tau_2\tau_3|T|\vec{\pi}, m'_{s2}m'_{s3}\tau'_2\tau'_3\rangle_a =$$
$$\langle \vec{p}, m_{s2}m_{s3}\tau_2\tau_3|T(1-P_{23})|\vec{\pi}, m'_{s2}m'_{s3}\tau'_2\tau'_3\rangle \quad (32)$$

is defined as the physical nucleon-nucleon $T$-matrix, where the anti-symmetrized NN basis states $|\vec{p}, m_{s2}m_{s3}\tau_2\tau_3\rangle_a$ explicitly include the individual nucleon's spin and isospin quantum numbers. In our calculation, we employ the nucleon-nucleon $T$-matrix in the momentum-helicity basis $|\vec{p}; \hat{\vec{p}} S\Lambda; t\rangle^{\pi a}$ where S, t, and $\Lambda$ denote the total spin, total isospin, and the helicity of the two-nucleon system, respectively. The superscript $\pi a$ indicates that the basis states have definite parity $\eta_\Pi$ and are anti-symmetrized. The physical $T$-matrix elements in Eq. (32) can be written in terms of the momentum-helicity representation $T_{\Lambda\Lambda'}^{\pi St}(\vec{p},\vec{\pi},E_p)$ [47]:

$$_a\langle \vec{p}, m_{s2}m_{s3}\tau_2\tau_3|T|\vec{\pi}, m'_{s2}m'_{s3}\tau'_2\tau'_3\rangle_a$$
$$= \frac{1}{4}\delta_{\tau_2+\tau_3,\tau'_2+\tau'_3} e^{-i(\Lambda_0\varphi_p - \Lambda'_0\varphi_\pi)} \sum_{S\Pi t}(1-\eta_\Pi(-)^{S+t})$$

*Contact author: r.ramazani@sci.ikiu.ac.ir

†Contact author: reza_ramazani@ut.ac.ir

$$C\left(\frac{1}{2}\frac{1}{2}t;\tau_2\,\tau_3\right)C\left(\frac{1}{2}\frac{1}{2}t;\tau'_2\tau'_3\right)C\left(\frac{1}{2}\frac{1}{2}S;m_{s2}m_{s3}\Lambda_0\right)$$
$$C\left(\frac{1}{2}\frac{1}{2}S;m'_{s2}m'_{s3}\Lambda'_0\right)\sum_{\Lambda\Lambda'} d_{\Lambda_0\Lambda}^S(\theta_p)d_{\Lambda'_0\Lambda'}^S(\theta_\pi)$$
$$T_{\Lambda\Lambda'}^{\pi St}(\vec{p},\vec{\pi},E_p). \quad (33)$$

In this expression, $d_{\Lambda\Lambda'}^S$ denotes the Wigner rotation matrix. By substituting Eq. (33) into Eq. (31), we obtain

$$U_1^{(1)}(\vec{p},\vec{q},\vec{q_0})$$
$$= \frac{1}{4}\sum_{m'_{s2}m'_{s3}\tau'_2\tau'_3} \int d\vec{q}''\, G_0^* \,\delta_{\tau_2+\tau_3,\tau'_2+\tau'_3}\,e^{-i(\Lambda_0\varphi_p-\Lambda'_0\varphi_\pi)}$$
$$\langle \vec{\pi}', m'_{s3}m_{s1}\tau'_3\tau_1|\langle \vec{q}'', m'_{s2}\tau'_2|T_F|\vec{q_0}, m_{s1}^0\tau_1^0\rangle$$
$$|\psi_d^{M_d}\rangle$$
$$\sum_{S\Pi t}(1-\eta_\Pi(-)^{S+t})\, C\left(\frac{1}{2}\frac{1}{2}t;\tau_2\,\tau_3\right)C\left(\frac{1}{2}\frac{1}{2}t;\tau'_2\tau'_3\right)$$
$$C\left(\frac{1}{2}\frac{1}{2}S;m_{s2}m_{s3}\Lambda_0\right)C\left(\frac{1}{2}\frac{1}{2}S;m'_{s2}m'_{s3}\Lambda'_0\right)$$
$$\sum_{\Lambda\Lambda'} d_{\Lambda_0\Lambda}^S(\theta_p)d_{\Lambda'_0\Lambda'}^S(\theta_\pi)T_{\Lambda\Lambda'}^{\pi St}(\vec{p},\vec{\pi},E_p). \quad (34)$$

Consider NN $T$-matrix elements $T_{\Lambda\Lambda'}^{\pi St}(\vec{p},\vec{p'},E_p)$ in the momentum-helicity representation. For nucleon-nucleon scattering calculations, it is convenient to orient the quantization axis (z axis) along the initial momentum vector $\vec{p}'$. As demonstrated in Ref. [45], the $T$-matrix $T_{\Lambda\Lambda'}^{\pi St}(\vec{p},\vec{p'},E_p)$ can then be rewritten as

$$T_{\Lambda\Lambda'}^{\pi St}(\vec{p},\vec{p'},E_p) =$$
$$\sum_{N=-S}^{S} \frac{e^{-iN(\varphi-\varphi')}d_{N\Lambda}^S(\theta)d_{N\Lambda'}^S(\theta')}{d_{\Lambda'\Lambda}^S(\theta'')} T_{\Lambda\Lambda'}^{\pi St}(p,p',\cos\theta'';E_p), \quad (35.a)$$

with

$$\cos\theta'' = \cos\theta'\cos\theta + \sin\theta'\sin\theta\cos(\varphi'-\varphi). \quad (35.b)$$

Substituting Eq. (35) into Eq. (34) yields

$$U_1^{(1)}(\vec{p},\vec{q},\vec{q_0}) = \frac{1}{4}\sum_{m'_{s2}m'_{s3}\tau'_2\tau'_3} \int d\vec{q}''$$
$$G_0^* \,\delta_{\tau_2+\tau_3,\tau'_2+\tau'_3}\,e^{-i(\Lambda_0\varphi_p-\Lambda'_0\varphi_\pi)}$$
$$\langle \vec{\pi}', m'_{s3}m_{s1}\tau'_3\tau_1|\langle \vec{q}'', m'_{s2}\tau'_2|T_F|\vec{q_0}, m_{s1}^0\tau_1^0\rangle$$
$$|\psi_d^{M_d}\rangle$$
$$\sum_{S\Pi t}(1-\eta_\Pi(-)^{S+t})\, C\left(\frac{1}{2}\frac{1}{2}t;\tau_2\,\tau_3\right)C\left(\frac{1}{2}\frac{1}{2}t;\tau'_2\tau'_3\right)$$
$$C\left(\frac{1}{2}\frac{1}{2}S;m_{s2}m_{s3}\Lambda_0\right)C\left(\frac{1}{2}\frac{1}{2}S;m'_{s2}m'_{s3}\Lambda'_0\right)$$

$$\sum_{\Lambda\Lambda'} d^S_{\Lambda_0\Lambda}(\theta_p) d^S_{\Lambda'_0\Lambda'}(\theta_\pi)$$

$$\sum_{N=-S}^{S} \frac{e^{-iN(\varphi_p-\varphi_\pi)} d^S_{N\Lambda}(\theta_p) d^S_{N\Lambda'}(\theta_\pi)}{d^S_{\Lambda'\Lambda}(\theta')} T^{\pi St}_{\Lambda\Lambda'}(p,\pi,\cos\theta';E_p), \quad (36)$$

where

$$\cos\theta' = \cos\theta_p \cos\theta_\pi + \sin\theta_p \sin\theta_\pi \cos(\varphi_p - \varphi_\pi). \quad (37)$$

Let us define

$$\mathcal{U}^{(1)}(\vec{p},\vec{q},\vec{\pi}) = \frac{1}{4}\delta_{\tau_2+\tau_3,\tau'_2+\tau'_3}$$
$$e^{-i(\Lambda_0\varphi_p - \Lambda'_0\varphi_\pi)}$$
$$\sum_{S\Pi t}(1-\eta_\Pi(-)^{S+t})\, C\left(\frac{1}{2}\frac{1}{2}t;\tau_2\,\tau_3\right) C\left(\frac{1}{2}\frac{1}{2}t;\tau'_2\tau'_3\right)$$
$$C\left(\frac{1}{2}\frac{1}{2}S;m_{s2}m_{s3}\Lambda_0\right) C\left(\frac{1}{2}\frac{1}{2}S;m'_{s2}m'_{s3}\Lambda'_0\right)$$
$$\sum_{\Lambda\Lambda'} d^S_{\Lambda_0\Lambda}(\theta_p) d^S_{\Lambda'_0\Lambda'}(\theta_\pi)$$
$$\sum_{N=-S}^{S} \frac{e^{-iN(\varphi_p-\varphi_\pi)} d^S_{N\Lambda}(\theta_p) d^S_{N\Lambda'}(\theta_\pi)}{d^S_{\Lambda'\Lambda}(\theta')} T^{\pi St}_{\Lambda\Lambda'}(p,\pi,\cos\theta';E_p), \quad (38)$$

and

$$T_F(\vec{\pi}',\vec{q}'',\overrightarrow{q_0}) =$$
$$<\vec{\pi}',m'_{s3}m_{s1}\tau'_3\tau_1|<\vec{q}'',m'_{s2}\tau'_2|T_F|\overrightarrow{q_0},m^0_{s1}\tau^0_1>$$
$$|\psi^{M_d}_d> \quad (39)$$

Applying Eqs. (38) and (39), one can rewrite Eq. (36) as

$$U^{(1)}_1(\vec{p},\vec{q},\overrightarrow{q_0}) = \sum_{m'_{s2}m'_{s3}\tau'_2\tau'_3} \int d\vec{q}''$$
$$\mathcal{U}^{(1)}(\vec{p},\vec{q},\vec{\pi}) G^*_0 T_F(\vec{\pi}',\vec{q}'',\overrightarrow{q_0}) \quad (40)$$

Substituting Eqs. (30) and (40) into Eq. (15) leads to

$$T_F(\vec{p},\vec{q},\overrightarrow{q_0}) = U^{(1)}_0(\vec{p},\vec{q},\overrightarrow{q_0}) + \sum_{m'_{s2}m'_{s3}\tau'_2\tau'_3} \int d\vec{q}''$$
$$\mathcal{U}^{(1)}(\vec{p},\vec{q},\vec{\pi}) \frac{m}{p^2+i\epsilon-\pi^2} T_F(\vec{\pi}',\vec{q}'',\overrightarrow{q_0}) \quad (41)$$

### III. A NEW APPROACH TO SINGULARITY TREATMENT

Two types of singularities arise in the nucleon-deuteron breakup channel, one originating from the free three-body propagator (moving singularity) and the other associated with the deuteron propagator. The latter introduces a pole in the two-body T-matrix at the

*Contact author: r.ramazani@sci.ikiu.ac.ir

†Contact author: reza_ramazani@ut.ac.ir

deuteron binding energy, $E_d$. The residue at this singularity is extracted explicitly by defining

$$\hat{T} = (E_P - E_d)T. \quad (42)$$

Such a singularity also appears in the operator $T_F$. Therefore, the corresponding residue is extracted using the definition

$$\hat{T}_F = (E_P - E_d)T_F. \quad (43)$$

Using the above definitions, the singularity associated with the two-body propagator in Eq. (41) is explicitly represented as

$$\hat{T}_F(\vec{p},\vec{q},\overrightarrow{q_0}) = \hat{U}^{(1)}_0(\vec{p},\vec{q},\overrightarrow{q_0}) +$$
$$\sum_{m'_{s2}m'_{s3}\tau'_2\tau'_3} \int d\vec{q}''\, \hat{\mathcal{U}}^{(1)}(\vec{p},\vec{q},\vec{\pi})$$
$$\frac{m}{p^2+i\epsilon-\pi^2} \frac{\hat{T}_F(\vec{\pi}',\vec{q}'',\overrightarrow{q_0})}{\frac{3}{4m}(q_0^2-q''^2)+E_d+i\epsilon-E_d} =$$
$$\hat{U}^{(1)}_0(\vec{p},\vec{q},\overrightarrow{q_0}) +$$
$$\sum_{m'_{s2}m'_{s3}\tau'_2\tau'_3} \int d\vec{q}''\, \hat{\mathcal{U}}^{(1)}(\vec{p},\vec{q},\vec{\pi})$$
$$\frac{m}{p^2+i\epsilon-\pi^2} \frac{4m\hat{T}_F(\vec{\pi}',\vec{q}'',\overrightarrow{q_0})}{3(q_0^2-q''^2)+i\epsilon} \quad (44)$$

Since $q_0 = \sqrt{\frac{4m}{3}(E-E_d)}$ and $E_d = -|E_d|$, it follows that $q_0 > \sqrt{\frac{4m}{3}E} = Q_{max} > \sqrt{mE} = Q$ for the Nd breakup channel. The free three-body singularity occurs at the momenta $q \leq Q_{max}$. The singularity of the deuteron propagator occurs at $q_0$, which therefore, lies outside the momentum domain relevant for the free three-body singularity. Consequently, the two-body (deuteron) pole can be treated separately for $q > Q_{max}$ using standard methods. For the case $q \leq Q_{max}$, the $q''$-domain is decomposed into $(0, Q_{max}) \cup (Q_{max}, \bar{q})$ where $\bar{q}$ is the cutoff of $q''$ integration. Splitting the integration over $q''$, Eq. (41) becomes

$$T_F(\vec{p},\vec{q},\overrightarrow{q_0}) = U^{(1)}_0(\vec{p},\vec{q},\overrightarrow{q_0})$$
$$+ \sum_{m'_{s2}m'_{s3}\tau'_2\tau'_3} \int_0^{Q_{max}} d\vec{q}''$$
$$\mathcal{U}^{(1)}(\vec{p},\vec{q},\vec{\pi}) \frac{m}{p^2+i\epsilon-\pi^2} T_F(\vec{\pi}',\vec{q}'',\overrightarrow{q_0})$$
$$+ \sum_{m'_{s2}m'_{s3}\tau'_2\tau'_3} \int_{Q_{max}}^{\bar{q}} d\vec{q}''$$

$$\mathcal{U}^{(1)}(\vec{p},\vec{q},\vec{\pi})\frac{m}{p^2-\pi^2}T_F(\vec{\pi}',\vec{q}'',\overrightarrow{q_0}) \quad (45)$$

To treat the second integration in Eq. (45), we need to explicitly identify the singularity of the two-body propagator, as described in Eq. (44). Therefore, Eq. (45) is rewritten as follows

$$\widehat{T}_F(\vec{p},\vec{q},\overrightarrow{q_0}) = \widehat{U}_0^{(1)}(\vec{p},\vec{q},\overrightarrow{q_0})$$
$$+ \sum_{m'_{s2}m'_{s3}\tau'_2\tau'_3}\int_0^{Q_{max}}d\vec{q}''$$
$$\widehat{\mathcal{U}}^{(1)}(\vec{p},\vec{q},\vec{\pi})\frac{m}{p^2+i\epsilon-\pi^2}\frac{4m\widehat{T}_F(\vec{\pi}',\vec{q}'',\overrightarrow{q_0})}{3(q_0^2-q''^2)}$$
$$+ \sum_{m'_{s2}m'_{s3}\tau'_2\tau'_3}\int_{Q_{max}}^{\bar{q}}d\vec{q}''$$
$$\widehat{\mathcal{U}}^{(1)}(\vec{p},\vec{q},\vec{\pi})\frac{m}{p^2-\pi^2}\frac{4m\widehat{T}_F(\vec{\pi}',\vec{q}'',\overrightarrow{q_0})}{3(q_0^2-q''^2)+i\epsilon} \quad (46)$$

The second integration in Eq. (46) is treated straightforwardly using the reduction method. Let us denote the second integration by $I_2$. To solve the first integration of Eq. (46), a novel approach is presented to treat the associated moving singularity.

The free three-body propagator generally depends on q, $q''$, and the angle between them, expressed as $x_{q''} = \hat{q}''.\hat{q}$. To simplify the treatment of the associated singularity, the free three-body propagator is rewritten in terms of the momentum p of the 23-subsystem and the variable $\pi$, yielding

$$G_0^* = \frac{m}{p^2+i\epsilon-\pi^2}. \quad (47)$$

Here, p and $\pi$ satisfy the Eq. (25) and Eq. (29), respectively. This reformulation allows the singularity to be treated using a reduction method described below. Considering the relation between $\vec{q}''$ and $\vec{\pi}$ in Eq. (28), one observes that $d\vec{q}'' = d\vec{\pi}$ for a fixed $\vec{q}$. Consequently, the first integration in Eq. (46) over $q''$ can be written in the form

$$\widehat{T}_F(\vec{p},\vec{q},\overrightarrow{q_0}) = \widehat{U}_0^{(1)}(\vec{p},\vec{q},\overrightarrow{q_0}) + I_2$$
$$+ \sum_{m'_{s2}m'_{s3}\tau'_2\tau'_3}\int d\vec{\pi}\frac{m}{p^2+i\epsilon-\pi^2}$$
$$\widehat{\mathcal{U}}^{(1)}(\vec{p},\vec{q},\vec{\pi})T_F\left(\left(-\frac{\vec{\pi}}{2}-\frac{3}{4}\vec{q}\right),\left(\vec{\pi}-\frac{\vec{q}}{2}\right),\overrightarrow{q_0}\right) \quad (48)$$

with $T_F\left(\left(-\frac{\vec{\pi}}{2}-\frac{3}{4}\vec{q}\right),\left(\vec{\pi}-\frac{\vec{q}}{2}\right),\overrightarrow{q_0}\right) \equiv \frac{4m\widehat{T}_F(\vec{\pi}',\vec{q}'',\overrightarrow{q_0})}{3(q_0^2-q''^2)}$.

Let us introduce β to collectively denote all discrete spin and isospin quantum numbers appearing in the summations of Eq. (48). In terms of this notation, Eq. (48) takes the form

*Contact author: r.ramazani@sci.ikiu.ac.ir

†Contact author: reza_ramazani@ut.ac.ir

$$\widehat{T}_F(\vec{p},\vec{q},\overrightarrow{q_0}) = \widehat{U}_0^{(1)}(\vec{p},\vec{q},\overrightarrow{q_0}) + I_2 + \sum_\beta$$
$$\int_{-1}^1 dx_\pi \int_{\frac{\bar{q}}{2}}^{\pi_{max}} d\pi \int_0^{2\pi} d\varphi_\pi$$
$$\frac{m\pi^2}{p^2+i\epsilon-\pi^2}\widehat{\mathcal{U}}^{(1)}(\vec{p},\vec{q},\vec{\pi})$$
$$T_{F,\beta}\left(\left(-\frac{\vec{\pi}}{2}-\frac{3}{4}\vec{q}\right),\left(\vec{\pi}-\frac{\vec{q}}{2}\right),\overrightarrow{q_0}\right), \quad (49)$$

where

$$\pi_{max} = \sqrt{Q_{max}^2 + \left(\frac{q}{2}\right)^2 - Q_{max}qx_{q''}}, \quad (50)$$

with $x_{q''} = \hat{q}''.\hat{q}$.

## A. AZIMUTHAL ANGLE INTEGRATIONS

Since the free three-body propagator does not depend on the azimuthal angle $\varphi_\pi$, the corresponding part of the integration can be performed independently. We therefore examine the dependence of each component of the integrand in Eq. (49), with particular focus on the variable $\varphi_\pi$. We choose a coordinate system in which the momentum q is aligned with the z-axis. The matrix element $T_{F,\beta}\left(\left(-\frac{\vec{\pi}}{2}-\frac{3}{4}\vec{q}\right),\left(\vec{\pi}-\frac{\vec{q}}{2}\right),\overrightarrow{q_0}\right)$ depends on the following quantities:

1. Magnitude of $|\vec{\pi}'|$:

$$|\vec{\pi}'| = \left|\frac{\vec{\pi}}{2}+\frac{3}{4}\vec{q}\right| = \sqrt{\left(\frac{\pi}{2}\right)^2 + \left(\frac{3}{4}q\right)^2 + \frac{3}{4}q\pi x_\pi}, \quad (51)$$

where $x_\pi = \hat{\pi}.\hat{q}$.

2. Angle between $\vec{\pi}'$ and $\overrightarrow{q_0}$:

$$y_{\pi'q_0} = \vec{\pi}'.\overrightarrow{q_0} = \frac{\frac{3}{4}q(\hat{q}.\widehat{q_0})+\frac{\pi}{2}(\hat{\pi}.\widehat{q_0})}{|\vec{\pi}'|} = \frac{\frac{3}{4}qx_q+\frac{\pi}{2}y_{\pi q_0}}{|\vec{\pi}'|}, \quad (52)$$

with

$$y_{\pi q_0} = x_{q_0}x_\pi + \sqrt{1-x_{q_0}^2}\sqrt{1-x_\pi^2}\cos(\varphi_{q_0}-\varphi_\pi). \quad (53)$$

3. Magnitude of $|\vec{q}''|$:

$$|\vec{q}''| = \left|\vec{\pi}-\frac{\vec{q}}{2}\right| = \sqrt{(\pi)^2 + \left(\frac{q}{2}\right)^2 - q\pi x_\pi}. \quad (54)$$

4. Angle between $(\vec{\pi}-\frac{\vec{q}}{2})$ and $\overrightarrow{q_0}$:

$$y_{\left(\vec{\pi}-\frac{\vec{q}}{2}\right)q_0} = (\vec{\pi}-\frac{\vec{q}}{2}).\overrightarrow{q_0} = \frac{-\frac{q}{2}(\hat{q}.\widehat{q_0})+\pi(\hat{\pi}.\widehat{q_0})}{\left|\vec{\pi}-\frac{\vec{q}}{2}\right|} = \frac{-\frac{q}{2}x_q+\pi y_{\pi q_0}}{\left|\vec{\pi}-\frac{\vec{q}}{2}\right|}, \quad (55)$$

with $y_{\pi q_0}$ given in Eq. (53),

5. Relative angle between $\vec{\pi}'$ and $\vec{q}''$:

$$x^{q_0}_{q''\pi'} =$$

$$x^{q_0}_{\left(\vec{\pi}-\frac{\vec{q}}{2}\right)\pi'} = \frac{\widehat{\left(\vec{\pi}-\frac{\vec{q}}{2}\right)}\cdot\widehat{\left(\frac{\vec{\pi}}{2}+\frac{3}{4}\vec{q}\right)} - [\widehat{(\vec{\pi}-\frac{\vec{q}}{2})}\cdot\widehat{q_0}][\widehat{(\frac{\vec{\pi}}{2}+\frac{3}{4}\vec{q})}\cdot\widehat{q_0}]}{\sqrt{1-[\widehat{(\vec{\pi}-\frac{\vec{q}}{2})}\cdot\widehat{q_0}]^2}\sqrt{1-[\widehat{(\frac{\vec{\pi}}{2}+\frac{3}{4}\vec{q})}\cdot\widehat{q_0}]^2}} =$$

$$\frac{\widehat{\left(\vec{\pi}-\frac{\vec{q}}{2}\right)}\cdot\widehat{\left(\frac{\vec{\pi}}{2}+\frac{3}{4}\vec{q}\right)} - y_{q''q_0}\, y_{\pi'q_0}}{\sqrt{1-(y_{q''q_0})^2}\sqrt{1-(y_{\pi'q_0})^2}},$$

(56)

where

$$\widehat{\left(\vec{\pi}-\frac{\vec{q}}{2}\right)}\cdot\widehat{\left(\frac{\vec{\pi}}{2}+\frac{3}{4}\vec{q}\right)} = \frac{\frac{\pi q}{2}x_\pi + \frac{\pi^2}{2} - \frac{3q^2}{4}}{\left|\vec{\pi}-\frac{\vec{q}}{2}\right|\left|\frac{\vec{\pi}}{2}+\frac{3}{4}\vec{q}\right|}.$$

(57)

Therefore, the matrix $T_{F,\beta}$ can be rewritten in terms of its relevant variables as

$$T_{F,\beta}\left(\left(-\frac{\vec{\pi}}{2}-\frac{3}{4}\vec{q}\right),\left(\vec{\pi}-\frac{\vec{q}}{2}\right),\vec{q_0}\right) =$$
$$T_{F,\beta}\left(|\vec{\pi}'|, y_{\pi'q_0}, \left|\vec{\pi}-\frac{\vec{q}}{2}\right|, y_{\left(\vec{\pi}-\frac{\vec{q}}{2}\right)q_0}, x^{q_0}_{\left(\vec{\pi}-\frac{\vec{q}}{2}\right)\pi'}\right),$$

(58)

where Eqs. (51)-(58) show that the matrix $T_{F,\beta}$ depends on $\cos(\varphi_{q_0}-\varphi_\pi)$. From the structure of Eq. (38) indicates that $\mathcal{U}^{(1)}(\vec{p},\vec{q},\vec{\pi})$ is proportional to $\cos(\Lambda_0\varphi_p - \Lambda'_0\varphi_\pi)$ with $\Lambda_0, \Lambda'_0 = 0,1$. When $\Lambda_0 = 0$, the expression becomes independent of $\varphi_\pi$. Furthermore, for $\Lambda'_0 = 0$, the integration over $\varphi_\pi$ becomes straightforward by fixing $\varphi_{q_0}$ to an arbitrary value (e.g., zero). In the case $\Lambda_0 = \Lambda'_0 = 1$, $\mathcal{U}^{(1)}(\vec{p},\vec{q},\vec{\pi})$ becomes proportional to $\cos(\varphi_p - \varphi_\pi)$. Since $\mathcal{U}^{(1)}(\vec{p},\vec{q},\vec{\pi})$ also contains the two-body $T$-matrix, $T^{\pi St}_{\Lambda\Lambda'}(p,\pi,\cos\theta';E_p)$, it generally depends on $\cos(\varphi_p - \varphi_\pi)$. Thus, the $\varphi_\pi$ dependence part of the integration in Eq. (49) takes the form

$$I_1(\varphi_{q_0},\varphi_p) = \int_0^{2\pi} F(\cos(\varphi_{q_0}-\varphi_\pi))G(\cos(\varphi_p-\varphi_\pi))d\varphi_\pi.$$

(59)

One finds that $I_1(\varphi_{q_0},\varphi_p) = I_1(\varphi_{q_0}-\varphi_p)$. Thus, for a chosen arbitrary value of $\varphi_{q_0}$, $\cos(\varphi_p)$ and $\sin(\varphi_p)$ are determined uniquely as [22]:

$$\cos\varphi_p = \cos\varphi_{q_0}\cos(\varphi_\pi-\varphi_{q_0}) - \sin\varphi_{q_0}\sin(\varphi_\pi-\varphi_{q_0}).$$

(60.a)

$$\sin\varphi_p = \sin\varphi_{q_0}\cos(\varphi_\pi-\varphi_{q_0}) - \cos\varphi_{q_0}\sin(\varphi_\pi-\varphi_{q_0}).$$

(60.b)

**B. SINGULARITY TREATMENT**

After performing the integration over $\varphi_\pi$, the kernel $\mathcal{U}^{(1)}$ in Eq. (38) reduces to

$$\mathcal{U}^{(1)}(\vec{p},\vec{q},\vec{\pi}) = \mathcal{U}^{(1)}(p,\hat{p}\cdot\hat{q}_0,\pi,\hat{\pi}\cdot\hat{q}_0,\hat{\pi}\cdot\hat{p})$$
$$= \frac{1}{4}\delta_{\tau_2+\tau_3,\tau'_2+\tau'_3}$$
$$\sum_{S\Pi t}(1-\eta_\Pi(-)^{S+t})\, C\left(\frac{1}{2}\frac{1}{2}t;\tau_2\,\tau_3\right)C\left(\frac{1}{2}\frac{1}{2}t;\tau'_2\tau'_3\right)$$
$$C\left(\frac{1}{2}\frac{1}{2}S;m_{s2}m_{s3}\Lambda_0\right)C\left(\frac{1}{2}\frac{1}{2}S;m'_{s2}m'_{s3}\Lambda'_0\right)$$
$$\sum_{\Lambda\Lambda'}d^S_{\Lambda_0\Lambda}(\theta_p)d^S_{\Lambda'_0\Lambda'}(\theta_\pi)$$
$$\sum_{N=-S}^{S}\frac{d^S_{N\Lambda}(\theta_p)d^S_{N\Lambda'}(\theta_\pi)}{d^S_{\Lambda'\Lambda}(\theta')}T^{\pi St}_{\Lambda\Lambda'}(p,\pi,\cos\theta';E_p). \quad (61)$$

By considering Eqs. (18) and (10), the quantities $U^{(1)}_0(\vec{p},\vec{q},\vec{q_0})$ and $T_F(\vec{p},\vec{q},\vec{q_0})$ can be represented as

$$U^{(1)}_0(\vec{p},\vec{q},\vec{q_0}) = U^{(1)}_0(p,\hat{p}\cdot\hat{q}_0,\pi_0,\widehat{\pi_0}\cdot\hat{q}_0,\widehat{\pi_0}\cdot\hat{p}) \equiv U^{(1)}_0(\zeta). \quad (62)$$

$$T_F(\vec{p},\vec{q},\vec{q_0}) = T_F(p,\hat{p}\cdot\hat{q}_0,\pi_0,\widehat{\pi_0}\cdot\hat{q}_0,\widehat{\pi_0}\cdot\hat{p}) \equiv T_F(\zeta). \quad (63)$$

where $\zeta \equiv p,\hat{p}\cdot\hat{q}_0,\pi_0,\widehat{\pi_0}\cdot\hat{q}_0,\widehat{\pi_0}\cdot\hat{p}$. Substituting Eqs. (58) and (61)-(63) into Eq. (49) leads to

$$\hat{T}_F(\zeta) = \hat{U}^{(1)}_0(\zeta) + I_2 + \sum_\beta \int_{-1}^{1}dx_\pi \int_{\frac{q}{2}}^{\pi_{max}}d\pi$$
$$\frac{m\pi^2}{p^2+i\varepsilon-\pi^2}\hat{\mathcal{U}}^{(1)}(p,\hat{p}\cdot\hat{q}_0,\pi,\hat{\pi}\cdot\hat{q}_0,\hat{\pi}\cdot\hat{p})T_{F,\beta}(\xi), \quad (64)$$

with the argument set $\xi$ given by

$$\xi \equiv |\vec{\pi}'|, y_{\pi'q_0},\left|\vec{\pi}-\frac{\vec{q}}{2}\right|, y_{\left(\vec{\pi}-\frac{\vec{q}}{2}\right)q_0}, x^{q_0}_{\left(\vec{\pi}-\frac{\vec{q}}{2}\right)\pi'}. \quad (65)$$

The singularity of the propagator is treated through the usual decomposition

$$\int_{k_{min}}^{k_{max}}\frac{k''F(k'')}{(k^2-k''^2+i\varepsilon)}dk''$$
$$= \mathcal{P}\int_{k_{min}}^{k_{max}}\frac{[k''F(k'')-k^2F(k)]}{(k^2-k''^2)}dk''$$
$$+ \frac{kF(k)}{2}\left(\ln\left(\frac{k-k_{min}}{k+k_{min}}\frac{k_{max}+k}{k_{max}-k}\right) - i\bar{\pi}\right)$$

(66)

where $\mathcal{P}$ denotes the principal value part and, the constant $\pi$ is denoted by $\bar{\pi}$ to avoid confusion with the momentum variable. Then, the integration is calculated as

$$\hat{T}_F(\zeta) = \hat{U}^{(1)}_0(\zeta) + I_2 + \sum_\beta \int_{-1}^{1}dx_\pi$$

*Contact author: r.ramazani@sci.ikiu.ac.ir

†Contact author: reza_ramazani@ut.ac.ir

$$\left\{ \mathcal{P} \int_{\frac{q}{2}}^{\pi_{max}} \frac{[\pi^2 F(\pi) - p^2 F(p)]}{(p^2 - \pi^2)} d\pi \right.$$

$$+ \frac{pF(p)}{2}\left( \ln\left(\frac{p - \frac{q}{2}}{p + \frac{q}{2}} \frac{\pi_{max} + p}{\pi_{max} - p}\right) \right.$$

$$\left.\left. - i\pi \right) \right\}$$

$$= \widehat{U}_0^{(1)}(\zeta) + I_2 + \sum_\beta \int_{-1}^{1} dx_\pi$$

$$\left\{ \mathcal{P} \int_{\frac{q}{2}}^{\pi_{max}} \frac{[\pi^2 F(\pi) - p^2 F(p)]}{(p^2 - \pi^2)} d\pi + \frac{pF(p)}{2}\left( \ln\left(\frac{p - \frac{q}{2}}{p + \frac{q}{2}}\right) + \ln\left(\frac{\pi_{max} + p}{\pi_{max} - p}\right) - i\pi \right) \right\}, \quad (67)$$

with

$$F(\pi) = m\widehat{U}^{(1)}\left(p, \widehat{p}.\widehat{q}_0, \pi, \widehat{\pi}.\widehat{q}_0, \widehat{\pi}.\widehat{p}\right)T_{F,\beta}(\xi), \quad (68)$$

and, $F(p) = F(\pi = p)$. Since $|x_{q''}| \leq 1$, the denominator of the second logarithm in Eq. (67) is zero only when both $x_{q''} = 1$ and $q = \frac{Q_{max}}{2}$ simultaneously. Therefore, for the case that $q = \frac{Q_{max}}{2}$ there is still a logarithmic singularity at $x_{q''} = 1$. In other words, the logarithmic singularity occurs at $(q, x_{q''}) = (\frac{Q_{max}}{2}, 1)$ that is treated using the standard methods such as the reduction method, see appendix A. Having $T_F$ in Eq. (67), one can construct the full Faddeev breakup amplitude, $U^{full}$. This amplitude is then used to evaluate the differential cross section for the nucleon–deuteron breakup process in the laboratory frame:

$$\frac{d^3\sigma}{dE_1 dk_1} = (2\pi)^4 \frac{m^3 p k_1}{2 k_{lab}} \frac{1}{6} \sum_{m_{s1} m_{s2} m_{s3} m_{s1}^0 M_d} \int dp |U^{full}|^2. \quad (69)$$

### IV. SUMMARY AND CONCLUSIONS

The earlier three-dimensional helicity-based formulation of the nucleon–deuteron breakup process accounted only for the leading-order term of the full Faddeev equation. In this work, we have extended that formalism by solving the complete Faddeev equation, omitting three-nucleon-force contributions. By treating momentum vectors directly rather than using partial-wave decomposition, the method naturally incorporates contributions from all partial waves, and therefore, the application of this formalism is not limited to any specific energy range of the scattering projectile.

*Contact author: r.ramazani@sci.ikiu.ac.ir

†Contact author: reza_ramazani@ut.ac.ir

Furthermore, we have presented, for the first time, a novel and systematic procedure for treating moving logarithmic singularities, analogous to the established handling of simple poles in the two-body T-matrix. This method eliminates the need to subdivide the q-domain, regarding the moving singularities, into multiple intervals to isolate the moving singularities. The key idea is to evaluate these singularities at the center-of-mass energy of the 23-subsytem and introduce an appropriate transformation that reduces the moving singularities to a form equivalent to a simple pole in a two-nucleon system. Evaluating three-nucleon singularities in a manner analogous to those of two-nucleon systems, without q-domain decomposition, substantially facilitates the numerical implementation of Faddeev scheme in few-body scattering problems.

The natural continuation of this work is the incorporation of three-nucleon-force (3NF) effects. Since the operator structure of typical 3NF models closely resembles that of the two-nucleon interaction, the present formalism can be extended to include 3NF contributions in a conceptually straightforward way. In addition, the suitability of the proposed singularity-handling scheme may be investigated in four-nucleon systems, with the objective of establishing a robust framework capable of mapping the more intricate singularity structures onto simpler, tractable forms.

### APPENDIX A

First, let us set $\epsilon$ near the singularity as

$$\epsilon = \pi_{max} - p. \quad (A)$$

The singular term is

$$\ln\left(\frac{\pi_{max} + p}{\pi_{max} - p}\right) = \ln\left(\frac{2p + \epsilon}{\epsilon}\right) = \ln\left(\frac{2p}{\epsilon}\right) + O(\epsilon). \quad (B)$$

When $\epsilon \to 0$, the kernel behaves as

$$\frac{1}{2} p F(p) \ln\left(\frac{1}{\epsilon}\right). \quad (C)$$

So, the singular part is

$$K_{sing}(x_{q''}) = \frac{1}{2} p F(p) \ln\left(\frac{1}{\epsilon}\right). \quad (D)$$

Near $x_{q''} = 1$, we expand $\pi_{max}$

$$\pi_{max}(x_{q''}) = \sqrt{Q_{max}^2 + \frac{q^2}{4} - Q_{max} q x_{q''}}. \quad (E)$$

At the singular point, $q = Q_{max}/2$

$$\pi_{max}(x_{q''}) = Q_{max}\sqrt{\frac{17}{16} - \frac{1}{2} x_{q''}}. \quad (F)$$

Differentiating as $x_{q''} = 1$

$$\frac{d\pi_{max}}{dx_{q''}} = -\frac{Q_{max}}{4\pi_{max}} = -\frac{Q_{max}}{4p}, \quad (G)$$

at the singular point. Thus, near the singularity, we have

$\epsilon = \pi_{max} - p \approx -\frac{Q_{max}}{4p}(x_{q''} - 1).$ (H)

Therefore:

$\ln\left(\frac{1}{\epsilon}\right) = \ln\left(\frac{4p}{Q_{max}}\right) + \ln\left(\frac{1}{1-x_{q''}}\right).$ (I)

Only the second part diverges. The reduction method splits the integral

$\int_{-1}^{1} K(x_{q''})dx_{q''} = \int_{-1}^{1}[K(x_{q''}) - K_{sing}(x_{q''})]dx_{q''} + \int_{-1}^{1} K_{sing}(x_{q''})dx_{q''}.$ (J)

The first term is regular and evaluated numerically and the second integral is performed analytically. Using H and I

$K_{sing}(x_{q''}) = \frac{1}{2}pF(p)\left[\ln\left(\frac{4p}{Q_{max}}\right) + \ln\left(\frac{1}{1-x_{q''}}\right)\right].$ (K)

Thus,

$\int_{-1}^{1} K_{sing}(x_{q''})dx_{q''} = \frac{1}{2}pF(p)\left[2\ln\left(\frac{4p}{Q_{max}}\right) + \int_{-1}^{1} \ln\left(\frac{1}{1-x_{q''}}\right)dx_{q''}\right].$ (L)

Computing the remaining integral

$\int_{-1}^{1} \ln(1-x)dx = 2(\ln 2 - 1).$ (M)

So,

$\int_{-1}^{1} \ln\left(\frac{1}{1-x}\right)dx = 2(1 - \ln 2).$ (N)

Thus, the singular integral becomes

$\int_{-1}^{1} K_{sing}(x_{q''})dx_{q''} = pF(p)\left[\ln\left(\frac{4p}{Q_{max}}\right) + 1 - \ln 2\right].$ (O)

Define the regularized kernel:

$K_{reg}(x_{q''}) = K(x_{q''}) - K_{sing}(x_{q''}).$ (P)

Then, the full integral is

$\int_{-1}^{1} K(x_{q''})dx_{q''} = \int_{-1}^{1} K_{reg}(x_{q''})dx_{q''} + pF(p)\left[\ln\left(\frac{4p}{Q_{max}}\right) + 1 - \ln 2\right].$ (Q)

All the singular behaviors have been removed, and only finite, smooth integrals remain.

*Contact author: r.ramazani@sci.ikiu.ac.ir

†Contact author: reza_ramazani@ut.ac.ir

*Contact author: r.ramazani@sci.ikiu.ac.ir

†Contact author: reza_ramazani@ut.ac.ir